\documentclass[12pt,a4paper]{article}
\usepackage[T2A]{fontenc}
\usepackage[utf8]{inputenc}
\usepackage[english]{babel}
\usepackage{amssymb}
\usepackage{amsmath}
\usepackage{graphicx}
\usepackage[small,sc,center]{titlesec}
\usepackage{indentfirst}
\usepackage{siunitx}
\usepackage[labelsep=period]{caption}
\usepackage{caption}
\begin{document}

\begin{center}
	\textbf{Evolutionary variations of superficial iron and calcium abundance in main sequence A stars}

	\vskip 3mm
	\textbf{Yu. A. Fadeyev${}^1$\footnote{E--mail: fadeyev@inasan.ru},
	        R. M. Bayazitov${}^1$}

\vskip 12pt
\textit{$^1$Institute of Astronomy, Russian Academy of Sciences, Moscow, 119017 Russia} \\

	Received May 25, 2026; revised July 11, 2026; accepted July 11, 2026
\end{center}

\vskip 10pt
\noindent

\textbf{Abstract} ---
Main sequence stellar evolution models were computed together with solution of the equations of atomic
diffusion for 16 elements from hydrogen to nickel.
The grid of evolutionary tracks comprises the models with stellar masses ranged from 1.4 to  $2.5M_\odot$
computed for initial helium and metal abundances $Y=0.28$ and $Z=0.02$, respectively.
The calculations were done for the mass loss rates
$10^{-15}M_\odot/\textrm{yr}\le\dot M\le 10^{-12}M_\odot/\textrm{yr}$ as well as for $\dot M=0$.
The high superficial abundance of iron in Am stars is shown to be due to the radiative acceleration
acting on the atoms of iron.
The significantly smaller absorption coefficient of calcium is responsible for its gravitational
settling and accumulation above its opacity maximum at $T\sim 10^6\:\textrm{K}$.
Recover of the superficial calcium abundance is due to plunge of the outer convection zone bottom
to layers with its excessive abundance.
A significant role in the evolutionary variations of superficial abundances of chemical elements
belongs to the intermediate convection zone arising for the first $\lesssim 300$ Myr due to
accumulation of the atoms of iron and nickel in the layers with temperature
$T\sim 2\times 10^5\:\textrm{K}$.
In stars with mass $M\lesssim 1.9M_\odot$ both the outer and intermediate convection zones merge
due to evolutionary descend of the bottom of the outer convection zone so that overabundant iron
and nickel are transported to the outer layers by convection.
The merging of the convective zones is responsible for considerable variations of superficial abundances
of calcium and iron with duration ranging from a quarter to a half of the main--sequence lifetime
depending on the stellar mass.
Therefore, Am stars as well as slowly rotating nonmagnetic A main--sequence stars have the
common origin, whereas appearance of their chemical anomalies depend of the stellar mass and age.


\newpage
\section{INTRODUCTION}

Main sequence stars of the spectral type A belong to the thin disk population of the Galaxy and
therefore formed from the protostellar gas clouds with slightly different chemical composition.
However spectroscopic observations of these relatively young stars show a great variety of
the chemical element abundances in their atmospheres.
Anomalies of chemical composition are detected not only in peculiar Ap stars with strong magnetic
fields (Preston 1974) but also in other A stars (Roman et al. 1948, Slettebak 1949, Cowley 1969).
In particular, Conti (1970) distinguished a group of Am stars with weak absorption lines of
calcium and scandium with simultaneously observed in their spectra strong iron lines.
At present the number of stars identified or suspected as those of the Am type is nearly 4300
(Renson and Manfroid 2009).

The superficial abundances of chemical elements in 120 nonmagnetic slowly rotating Am and Fm stars
were presented by Ghazaryan et al. (2018) who collected the published data based
on the high resolution spectroscopy.
Unfortunately, these abundances were obtained in LTE approximation and sometimes with different
oscillator strengths for the same spectral lines.
One should be noted that non--LTE corrections applied in determination of elemental abundances
might be as high as 0.40 dex for the line Ca~I~4226\AA\ and 0.48 dex for the line Sc II 4246\AA,
so that Mashonkina and Fadeyev (2024) published the list of 54 Am stars with calcium and
scandium abundances determined on the basis of the high resolution spectroscopy and homogeneous
procedure of non--LTE calculations.

Abundance anomalies in the atmospheres of Am stars are believed to be due to diffusion separation
of chemical elements arising from different velocities of gravitational settling and radiative levitation
of various atoms of the stellar matter (Michaud 1970; Michaud et al. 1976; Vauclair et al. 1978).
More detailed computations
(Richer et al. 1992; 1993; 2000; Turcotte et al. 1998; Vick et al. 2010; Campilho et al. 2022;
Moedas et al. 2022; Hui--Bon--Hoa et al. 2022; Mashonkina and Fadeyev 2024; Hui--Bon--Hoa 2024)
showed that abundances of chemical elements in Am stars generally agree with theoretical
predictions of the diffusion model developed by Michaud (1970).
Nevertheless, theoretical abundances of the iron group elements are found to be often overestimated.
Disagreement between the theory and observations seems to be due to underestimation of mass loss
effects (Vick et al. 2010; Michaud et al. 2011; Hui--Bon--Hoa 2022).
One cannot also exclude effects of stellar matter mixing due to stellar rotation
(Deal et al. 2020; Dumont et al. 2021).

One of essential problems in the theory of Am stars remaining so far unsolved is the
origin of significant difference in abundances of chemical elements in the main sequence stars
with close values of their age and mass.
Explanation of these anomalies is the primary goal of the present study.
To this end we solve the equations of stellar evolution together with those of atomic
diffusion describing gravitational settling and radiative acceleration.
Results of calculations are compared with estimates of elemental abundances based on the
high resolution spectroscopy and non--LTE model atmosphere calculations.

\section{METHOD OF COMPUTATION}

In the present study we consider stellar evolution from the gravitational contraction of the
protostellar cloud up to almost complete exhaustion of hydrogen in the stellar center when its
central abundance becomes as low as $X_\mathrm{H,c}=10^{-4}$.
Computations were done with initial mass fractional abundances of helium $Y=0.28$ and heavier
elements $Z=0.02$ with elemental metal proportions derived by Grevesse and Sauval (1998).
Axial stellar rotation was assumed to be absent.

To calculate the stellar evolution we used the code MESA revision 24.08.1 (Paxton et al. 2018).
The energy generation rates and equations of nucleosynthesis were calculated for the grid
consisting of 19 isotopes from hydrogen ${^1}\textrm{H}$ to cuprum ${}^{59}\textrm{Cu}$ and
31 reactions.
Convective mixing was treated according to B\"ohm-Vitense (1958) with the mixing length to pressure
scale height ratio $\alpha_\mathrm{MLT}=1.8$.
In the layers where the Ledoux criterion is fulfilled we employed the approximation of semiconvection
with parameter $\alpha=0.1$ (see, for example, Langer et al. 1985).
Extra mixing at edges of convection zones was treated according to Herwig (2000) with
exponential overshooting parameter $f_\mathrm{ov}=0.014$.

Integration of the equations of atomic diffusion (Burgers, 1969) was done for all 19 isotopes of
the nuclear network at every time step of the solution of of stellar evolution equations.
To this end we employed the method by Thoul et al. (1994) extended by taking into account the radiative
acceleration (Hu et al. 2011).
Radiative forces were computed using the data of monochromatic absorption coefficients for 17 elements
(Seaton 2005) and in the present study effects of radiative levitation were considered for following
16 elements:
${}^1\textrm{H}$,     ${}^4\textrm{He}$,     ${}^{12}\textrm{C}$,   ${}^{14}\textrm{N}$,
${}^{16}\textrm{O}$,  ${}^{20}\textrm{Ne}$,  ${}^{23}\textrm{Na}$,  ${}^{24}\textrm{Mg}$,
${}^{27}\textrm{Al}$, ${}^{28}\textrm{Si}$,  ${}^{32}\textrm{S}$,   ${}^{40}\textrm{Ca}$,
${}^{52}\textrm{Cr}$, ${}^{55}\textrm{Mn}$,  ${}^{56}\textrm{Fe}$ and ${}^{58}\textrm{Ni}$.
Initial conditions for the consistent calculation of stellar evolution and atomic diffusion were
determined at the pre--main sequence stage after appearance of the radiative zone with
temperature $T\sim 3\times 10^6\:\textrm{K}$ in the interiors of the contracting stars.

Calculation of the evolution of A stars together with solution of the equations of atomic diffusion
should be done under the assumption of existence of the stellar wind since otherwise the atoms
of the iron group elements become overabundant at the outer boundary of the stellar model.
In the present study these are chromium, manganese, iron and nickel.
Unfortunately, observational estimates of mass loss rates in A stars are still highly uncertain.
In our study the upper limit of mass loss rate was assumed to be $\dot M\sim 10^{-12}M_\odot/\textrm{yr}$.
This estimate was obtained by Bertin et al. (1995) from measurements of blue wings of Mg~II and H~I
resonance lines of Sirius A.
Thus, for each value of the stellar mass $M$ we carried out the evolution calculations
both without the stellar wind ($\dot M=0$) and with several values of the mass loss rate
($10^{-15}M_\odot/\textrm{yr}\le \dot M\le 10^{-12}M_\odot/\textrm{yr}$) which was assumed to be
constant during the entire main sequence stage.
Below we show that this method allows us to get the rough order of magnitude estimate of $\dot M$
corresponding to the best agreement between the theory and observations.

\section{RESULTS OF COMPUTATIONS}

Let us set the star age $t_\mathrm{ev}$ to zero at cessation of gravitational contraction
and the onset of the steady--state hydrogen nuclear burning so that during the phase of
gravitational contraction $t_\mathrm{ev} < 0$.
For the sake of graphical representation we consider below the elemental abundances inside
the star $X$ and at the outer boundary $X_\mathrm{s}$ expressed in units of the initial value
$X_0$.

Figs.~\ref{fig1} and \ref{fig2} show variations of the surface abundance of calcium and iron in the
models of evolutionary sequences with mass $M=1.5$ and $2M_\odot$ for mass loss rates
$\dot M = 10^{-12}$, $10^{-13}$, $10^{-14}$ and $10^{-15}M_\odot/\textrm{yr}$.
Results of computations without mass loss ($\dot M = 0$) are shown by dotted lines.
The plots in Figs.~\ref{fig1} and \ref{fig2} encompass the time interval $t_\mathrm{ev}$ where
stellar evolution was calculated together with atomic diffusion.

\subsection{\textit{Early evolutionary phase}}

A common feature of dependencies similar to those shown in Figs.~\ref{fig1} and \ref{fig2}
is the rapid increase of the superficial calcium abundance at $t_\mathrm{ev} \approx 0$
because of radiative expulsion and displacement of the convection zone bottom to the photosphere.
This feature is illustrated in Fig.~\ref{fig3}, where the plots of the radial distribution of
calcium abundance and the ratio of radiative expulsion to the gravity are shown for the model
with mass $M=2M_\odot$ within the stellar age interval $-3\:\textrm{Myr}\le t_\mathrm{ev}\le 12.7\:\textrm{Myr}$.
The temperature of the layers with maximum absorption by atoms of calcium is
$T\approx 1.2\times 10^5\:\textrm{K}$.
Displacement of the bottom of the outer convection zone above these layers is accompanied by
concentration growth of calcium atoms in the stellar atmosphere because of their radiative
expulsion from the zone with maximum radiative acceleration.
It is interesting to note that the evolutionary variations of the superficial abundance of
scandium take place in a similar way during the early evolution phase
(see, e.g., Hui--Bon--Hoa et al. 2022) since the maxima of calcium and scandium absorption
coefficients locate in stellar interiors near one another because of small difference in their
ionization potentials.

As seen in Fig.~\ref{fig3}, the vertical stratification of chemical elements is not present
in the outer layers with temperature $T\lesssim 5\times 10^4\:\textrm{K}$ and this is due
not only to the mixing of stellar matter in the outer convection zone.
In the outermost layers with radiative transfer ($T\lesssim 10^4\:\textrm{K}$)
the vertical stratification cannot appear because of the small length of the stellar atmosphere
(in radius $\lesssim 10^{-4}R$, in mass $\lesssim 10^{-12}M$) as well as because of the high velocity
of diffusion due to the low gas density.

\subsection{\textit{Formation of the intermediate convection zone}}

\begin{figure}
 \centering
 \includegraphics[width=0.45\columnwidth,clip]{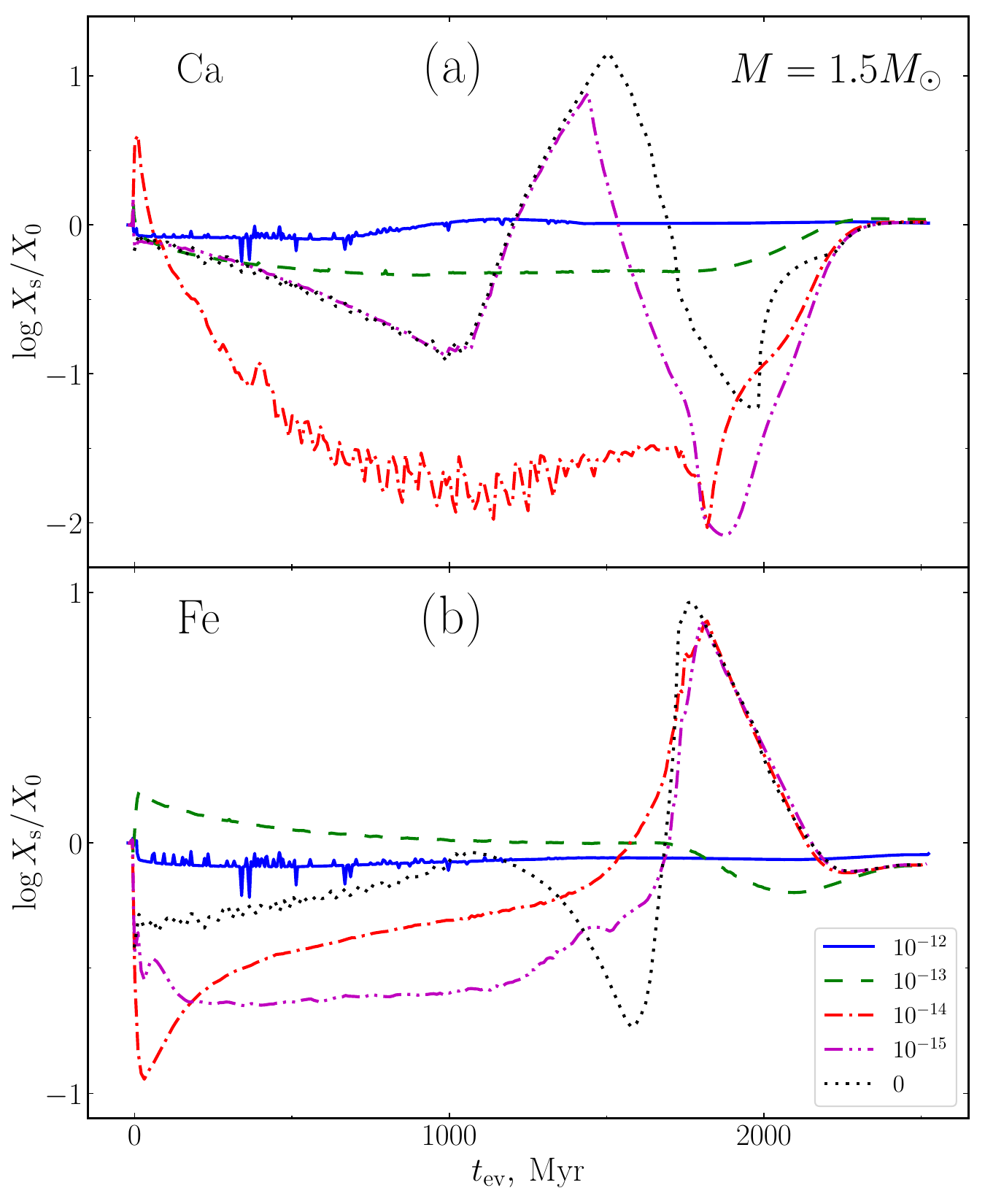}
 \caption{Evolutionary variations of the superficial abundance of calcium (a) and iron(b) in the
          models with mass $M=1.5M_\odot$ and mass loss rates
          $\dot M = 10^{-12}$ (solid lines), $10^{-13}$ (dashed lines),
          $10^{-14}$ (dash--dotted lines), $10^{-15} M_\odot/\textrm{yr}$ (dash--double dotted lines).
          Results of computations without mass loss ($\dot M = 0$) are shown by dotted lines.}
\label{fig1}
\end{figure}

\begin{figure}
 \centering
 \includegraphics[width=0.45\columnwidth,clip]{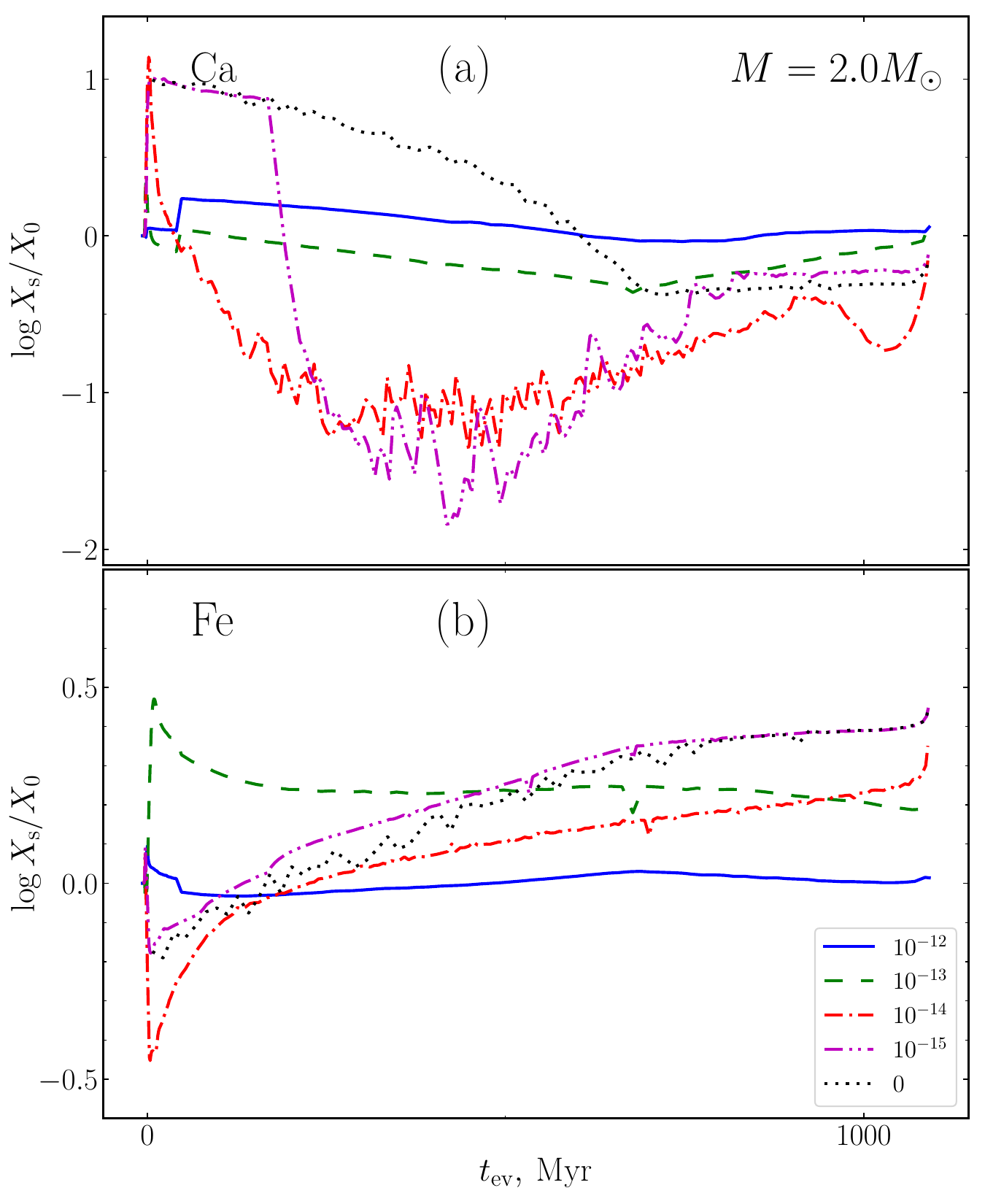}
 \caption{Same as Fig.~\ref{fig1} but for models of stars with mass $M=2M_\odot$.}
\label{fig2}
\end{figure}

\begin{figure}
 \centering
 \includegraphics[width=0.6\columnwidth,clip]{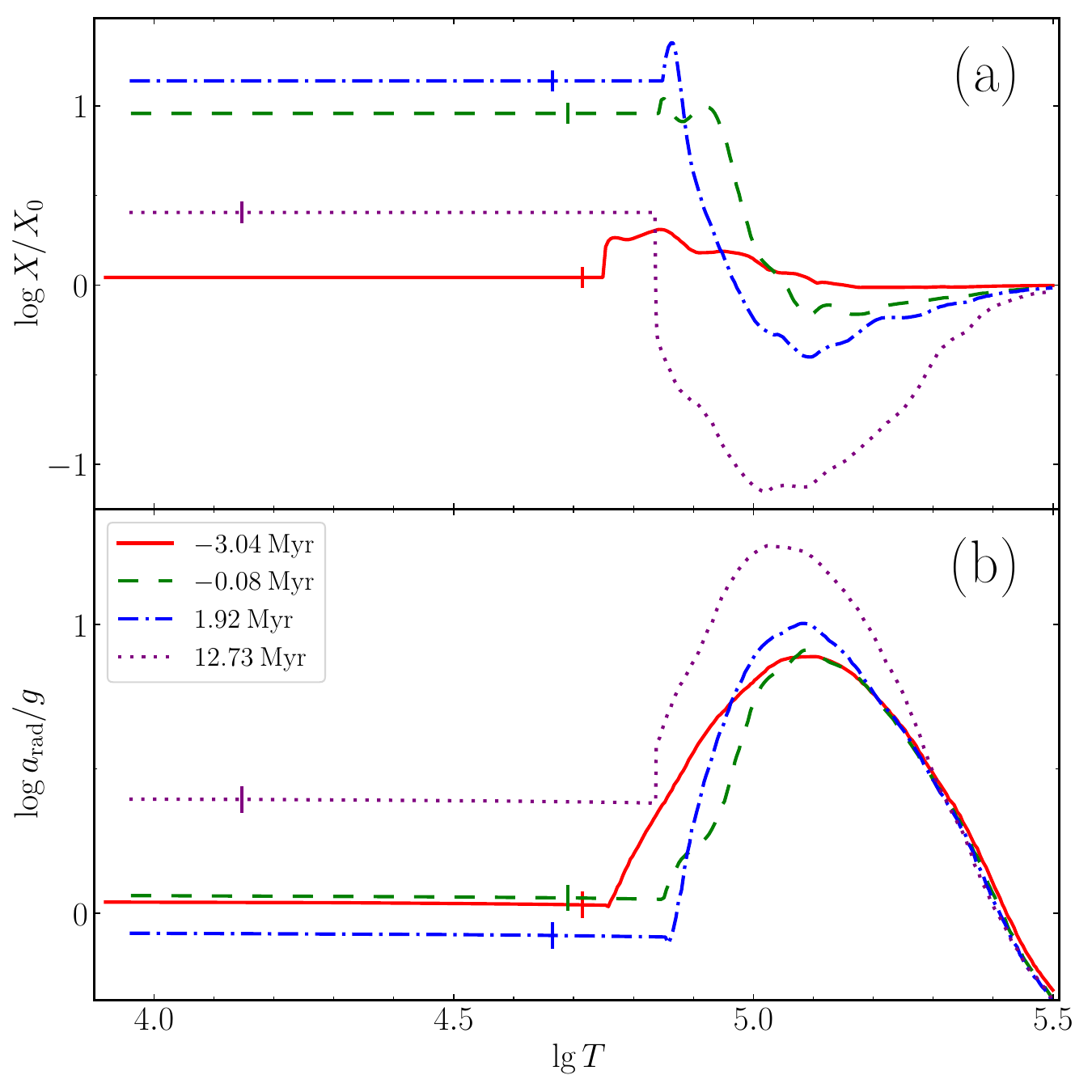}
 \caption{Radial distribution of calcium abundance (a) and the ratio of radiative acceleration
          to gravity (b) in outer layers of the star with mass $M=2M_\odot$ and the mass loss rate
          $\dot M=10^{-14}M_\odot/\textrm{yr}$ at the stellar age
          $t_\mathrm{ev}=-3.04$, $-0.08$, $1.92$ and $12.73\:\textrm{Myr}$.
          The bottom of the outer convection zone is marked by vertical dashes on the curves.}
\label{fig3}
\end{figure}

Returning to the plots in Figs.~\ref{fig1} and \ref{fig2}, we have to note that evolutionary
sequences with mass loss rates $\dot M\ge 10^{-13}M_\odot/\textrm{yr}$ do not show
significant variations of calcium and iron abundance during the whole stage of evolution.
This is due to the fact that the high rate of gas outflow from the stellar surface eliminates
appearance of anomalies of chemical composition because of significantly lower rates of
atomic diffusion.
Therefore, to explain chemical anomalies in Am stars we should consider the evolutionary
sequences computed with mass loss rates $\dot M\le 10^{-14}M_\odot/\textrm{yr}$.

As seen in Figs.~\ref{fig1} and \ref{fig2}, the ranges of evolutionary variations of superficial
calcium and iron abundance in models with $\dot M\le 10^{-14}M_\odot/\textrm{yr}$ are comparable
with abundance scatter of calcium and iron observed in Am stars:
$0.08\lesssim [\textrm{Fe}/\textrm{H}]\lesssim 0.71$,
$-0.87\lesssim [\textrm{Ca}/\textrm{H}]\lesssim 0.57$ (Mashonkina and Fadeyev 2024).
However first of all we have to note significant differences in the shape of plots
in Figs.~\ref{fig1} and \ref{fig2}.
In particular, in models with $M\approx 1.5M_\odot$ significant variations of the surface
abundance $X_\mathrm{s}$ go on from 20\% to 40\% of the main sequence lifetime, whereas
in models with $M=2M_\odot$ evolutionary variations of $X_\mathrm{s}$ remain almost
monotonous for the major fraction of the main sequence stage.
To explain such a difference we have to consider evolutionary variations of spatial abundance
distribution of the elements of the iron group during the first two or three hundred million years.

\begin{figure}
 \centering
 \includegraphics[width=0.6\columnwidth,clip]{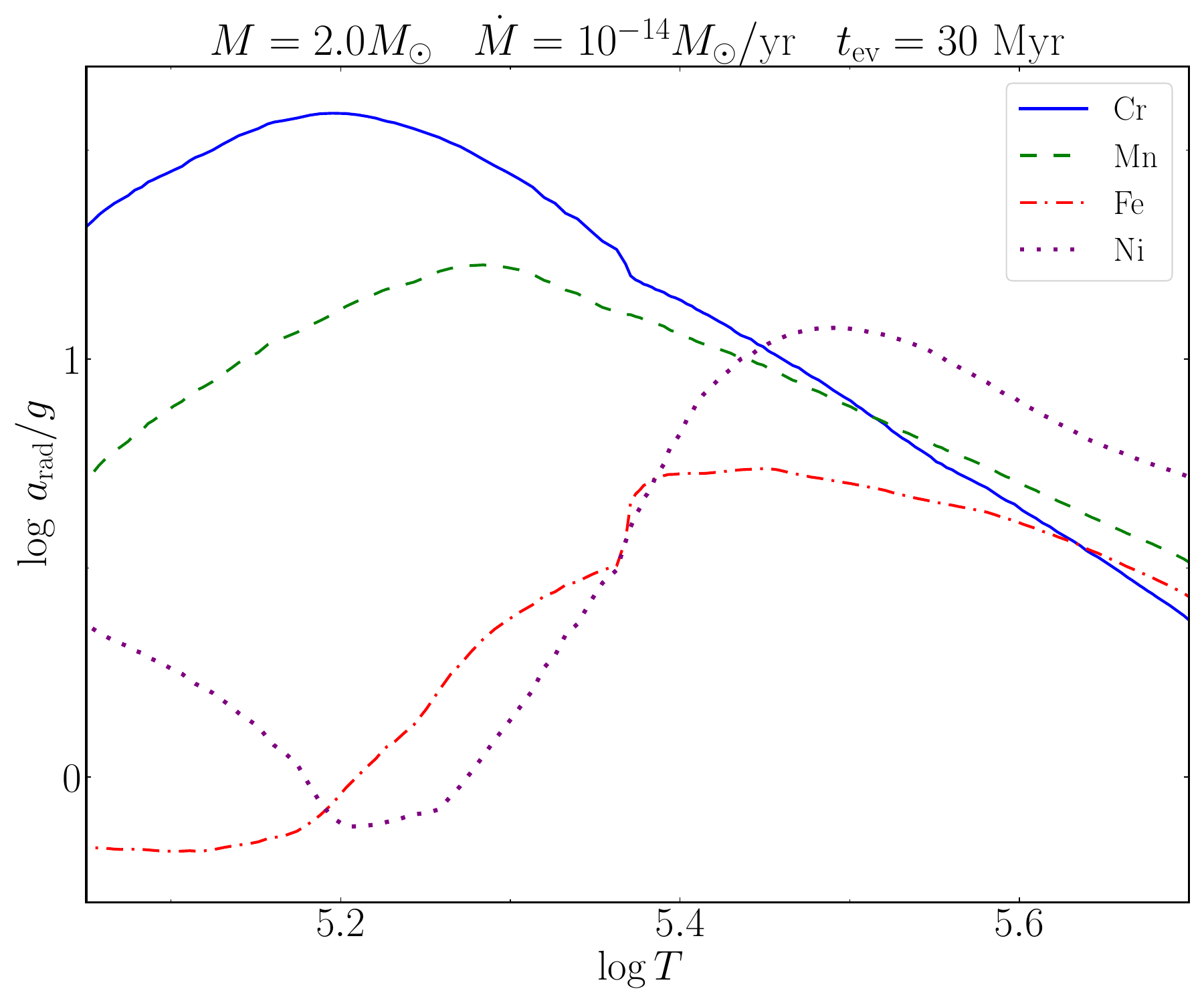}
 \caption{Radiative acceleration to gravity ratio for atoms of chromium, manganese, iron and nickel
          in the evolutionary model $M=2M_\odot$, $\dot M = 10^{-14}M_\odot/\textrm{yr}$,
          $t_\mathrm{ev}=30\ \textrm{Myr}$.}
\label{fig4}
\end{figure}

In the layers with gas temperature $T\lesssim 10^6\:\textrm{K}$ absorption coefficients of iron
and nickel increase with increasing radius (i.e. with decreasing $T$), reach their maxima at
$T\approx 2\times 10^5\:\textrm{K}$ and then decrease outward.
Therefore, the maximal diffusion velocity in vicinity of the opacity maximum
(the so called Z--bump, Iglesias et al. 1987; Seaton and Badnell 2004)
becomes responsible for accumulation of elements with highest absorption coefficient.
However, among four elements of the iron group (chromium, manganese, iron and nickel)
considered in the present study only two were found to accumulate at the Z--bump.
Absence of this feature in chromium and manganese is due to the lower temperature
corresponding to the maximum of the radiative acceleration (see Fig.~\ref{fig4}).

\begin{figure}
 \centering
 \includegraphics[width=0.6\columnwidth,clip]{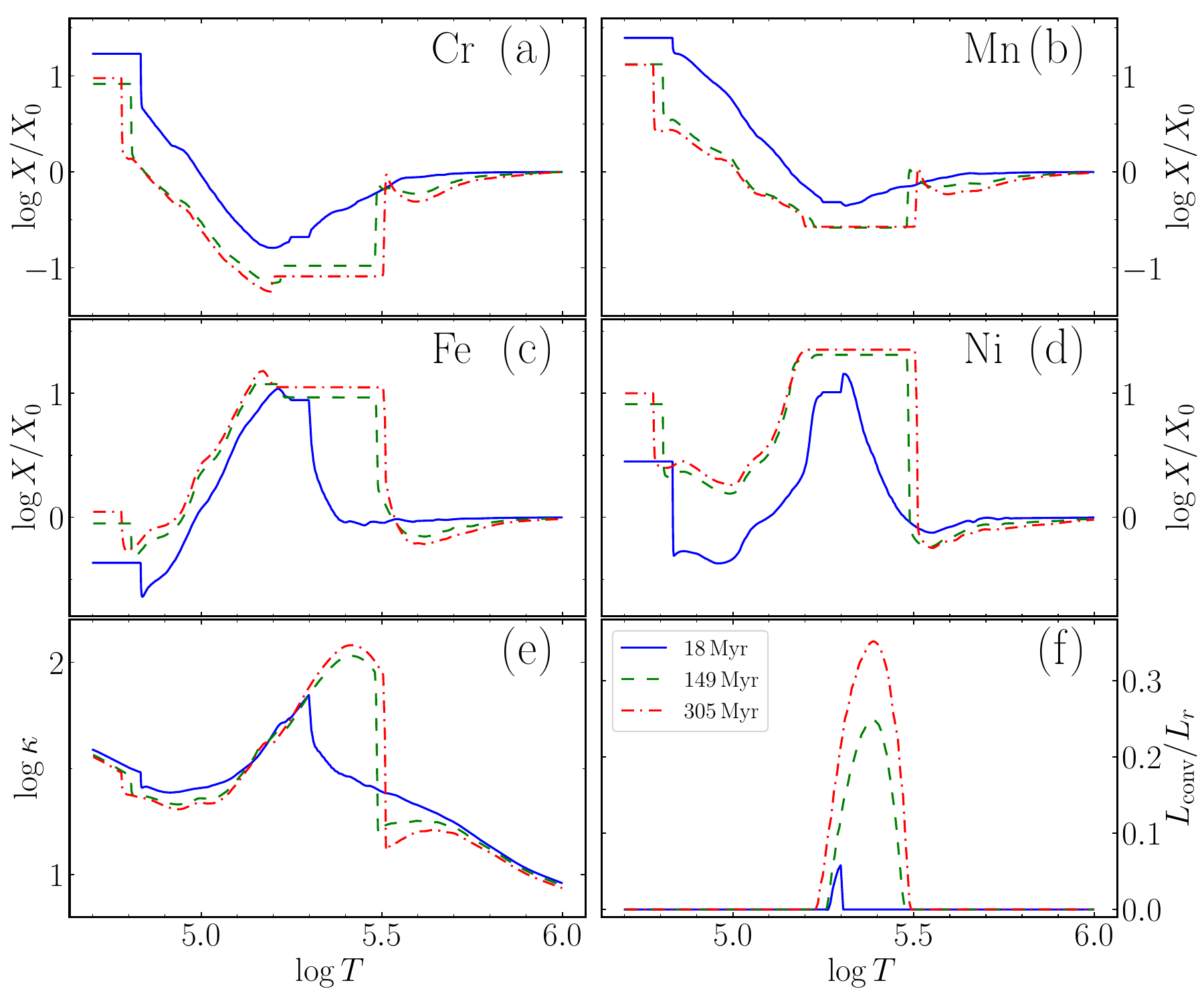}
 \caption{Radial abundance distributions of chromium (a), manganese (b), iron (c), nickel (d),
          the Rosseland mean opacity coefficient (e) and
          the fractional convection luminosity (f) in the star with mass $M=2M_\odot$ and
          the mass loss rate $\dot M=10^{-14}M_\odot/\textrm{yr}$.
          The plots correspond to the following values of the stellar age:
          $t_\mathrm{ev}=18$ (solid lines), $149$ (dashed lines) and
          $305\:\textrm{Myr}$ (dash--dotted lines).}
\label{fig5}
\end{figure}

In Fig.~\ref{fig5} we plotted the relative abundance of chromium, manganese, iron and nickel
as well as the Rosseland opacity coefficient and the ratio $L_\mathrm{conv}/L_r$ as a function $\log T$
for the three models of the evolutionary sequence $M=2M_\odot$, $\dot M = 10^{-14}M_\odot/\mathrm{yr}$
with the stellar age $t_\mathrm{ev}=18$, 149 and 305~Myr.
Here $L_\mathrm{conv}$ is the convective luminosity and $L_r$ is the total luminosity.
The growth of the iron and nickel abundances near the opacity maximum occurs until the gas layers
become convectively unstable and convective mixing leads to the constant spatial distribution
of elemental abundances.

The intermediate convection zone located between the convective core and the outer convection zone
appears due diffusive element separation and does not exist in the stellar evolution models
calculated without accounting for the effects of atomic diffusion.
The zone of the turbulent mixing at the temperature $T\sim 1\times 10^5\:\textrm{K}$ was mentioned
in (Richer et al. 2000; Vick et al. 2010; Moedas et al. 2022; Hui--Bon--Hoa et al. 2022;
Hui--Bon--Hoa 2024) but the role of this zone was taken into account using the simplified method
on the basis of the turbulent parameter $\omega$ (Richer et al. 2000).
In the present study both the convective heat flow and convective mixing were treated with the mixing
length theory (B\"ohm-Vitense 1958) including the solution of the diffusion equation for convective
elements in the whole stellar model (see, for example, Langer et al. 1985).
It should also be noted that the intermediate convection zone appears in stars with mass
$M\ge 1.4M_\odot$ since at lower stellar mass the bottom of the outer convection zone locates
deeper than the layers with temperature $2\times 10^5\:\textrm{K}$.
In the stellar evolution models computed with $Z=0.02$ the fraction of the energy transported by
the convective luminosity within the intermediate convection zone varies from $\approx 10\%$ for
$M=1.4M_\odot$ to $\approx 40\%$ for $M=2.5M_\odot$.

\begin{figure}
 \centering
 \includegraphics[width=0.6\columnwidth,clip]{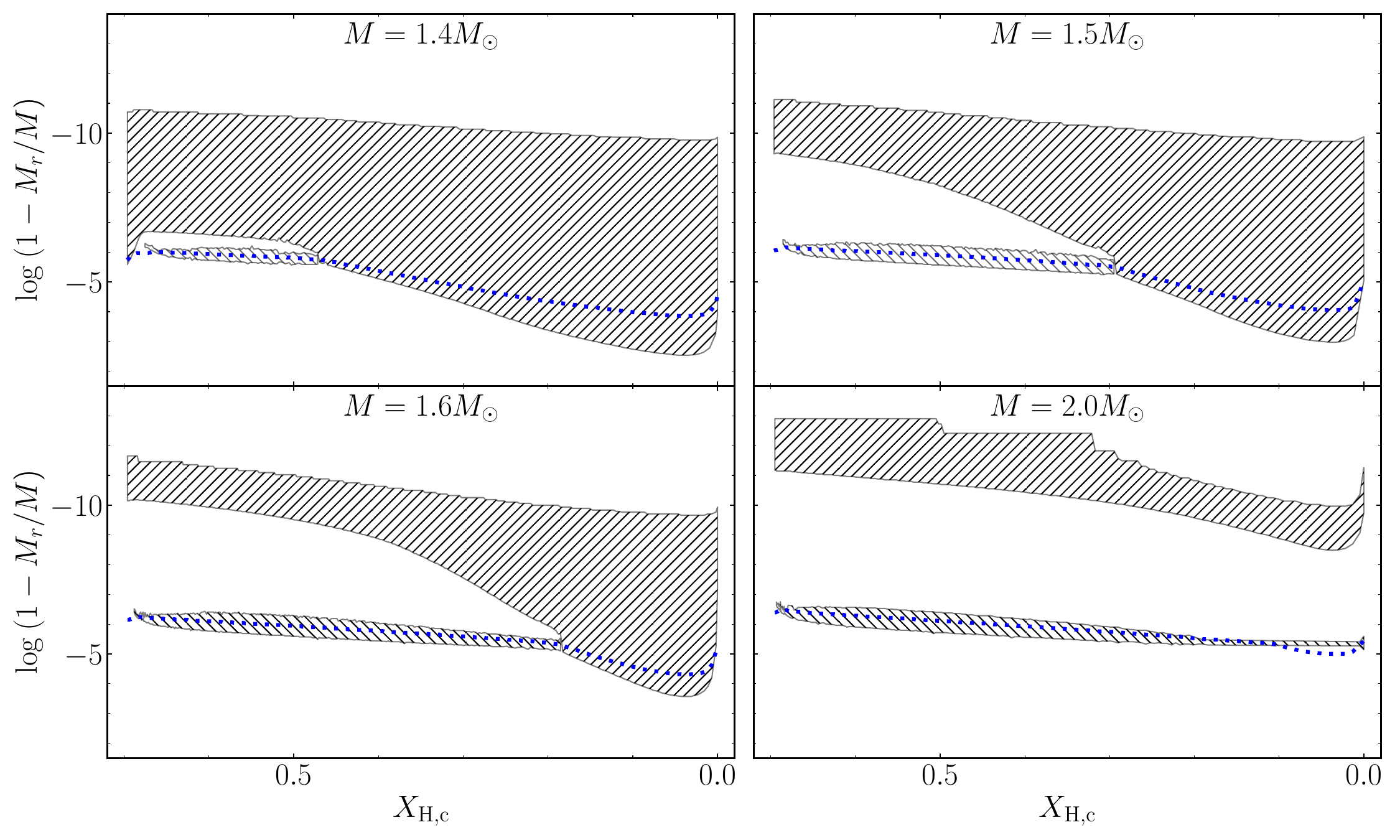}
 \caption{Lagrangean coordinates of the boundaries of outer and intermediate convection zones
          as a function of central hydrogen abundance $X_\mathrm{H,c}$ in the evolutionary
          sequences $M=1.4$, 1.5, 1.6 and $2M_\odot$ with $\dot M=10^{-14}M_\odot/\textrm{yr}$.
          Dotted lines indicate the layer with temperature $2.2\times 10^5\:\textrm{K}$.}
\label{fig6}
\end{figure}

Fig.~\ref{fig6} shows the Kippenhahn diagrams of evolutionary variations of the edges of the
outer and intermediate convection zones in the stellar models of evolutionary sequences
$M=1.4$, 1.5, 1.6 and $2M_\odot$ with mass loss rate $\dot M=10^{-14}M_\odot$.
For the sake of convenience the diagrams are plotted as a function of central hydrogen abundance
$X_\mathrm{H,c}$ whereas the dimensionless Lagrangean coordinate $1-M_r/M$ is plotted along the
vertical axis.
Here $M_r$ is the mass confined inside the spherical volume of radius $r$.
Both the outer and intermediate convection zones are represented by dashed areas.

As seen in Fig.~\ref{fig6}, the mass of the outer convection zone monotonously increases
because of plunging bottom of the outer convection zone whereas the edges of the
intermediate convection zone are almost unchanged.
Of great interest are the stars with mass $1.4M_\odot\le M\le 1.8M_\odot$ where
the outer and intermediate convection zones merge during the evolution.
The number density of iron and nickel in the intermediate convection zone is significantly higher
than that in the outer convection zone so that mixing of matter after merge of convection zones
leads to substantial enhancement of the superficial abundance of iron and nickel.
For example, in the models of the evolutionary sequence $M=1.5M_\odot$,
$\dot M\le 10^{-14}M_\odot/\textrm{yr}$ the convection zones merge at the star age
$t_\mathrm{ev}\approx 1.8\times 10^3\:\textrm{Myr}$ corresponding to the central hydrogen
abundance $X_\mathrm{H,c}\approx 0.3$ (see Fig.~\ref{fig6}) and the superficial iron abundance
increases by an order of magnitude (see Fig.~\ref{fig1}b).

However, one should bear in mind that the Kippenhahn diagrams in Fig.~\ref{fig6} give
only schematic representation of evolutionary variations of edges of convective zones
that are indistinct due to overshooting effects.
Overshooting and the limited velocity of convection are responsible for
substantial duration of temporal variations of the superficial abundance of chemical elements
which lasts from one quarter to one third of the main sequence lifetime.

\subsection{\textit{Variations of iron and calcium superficial abundance during the main sequence stage}}

\begin{figure}
 \centering
 \includegraphics[width=0.6\columnwidth,clip]{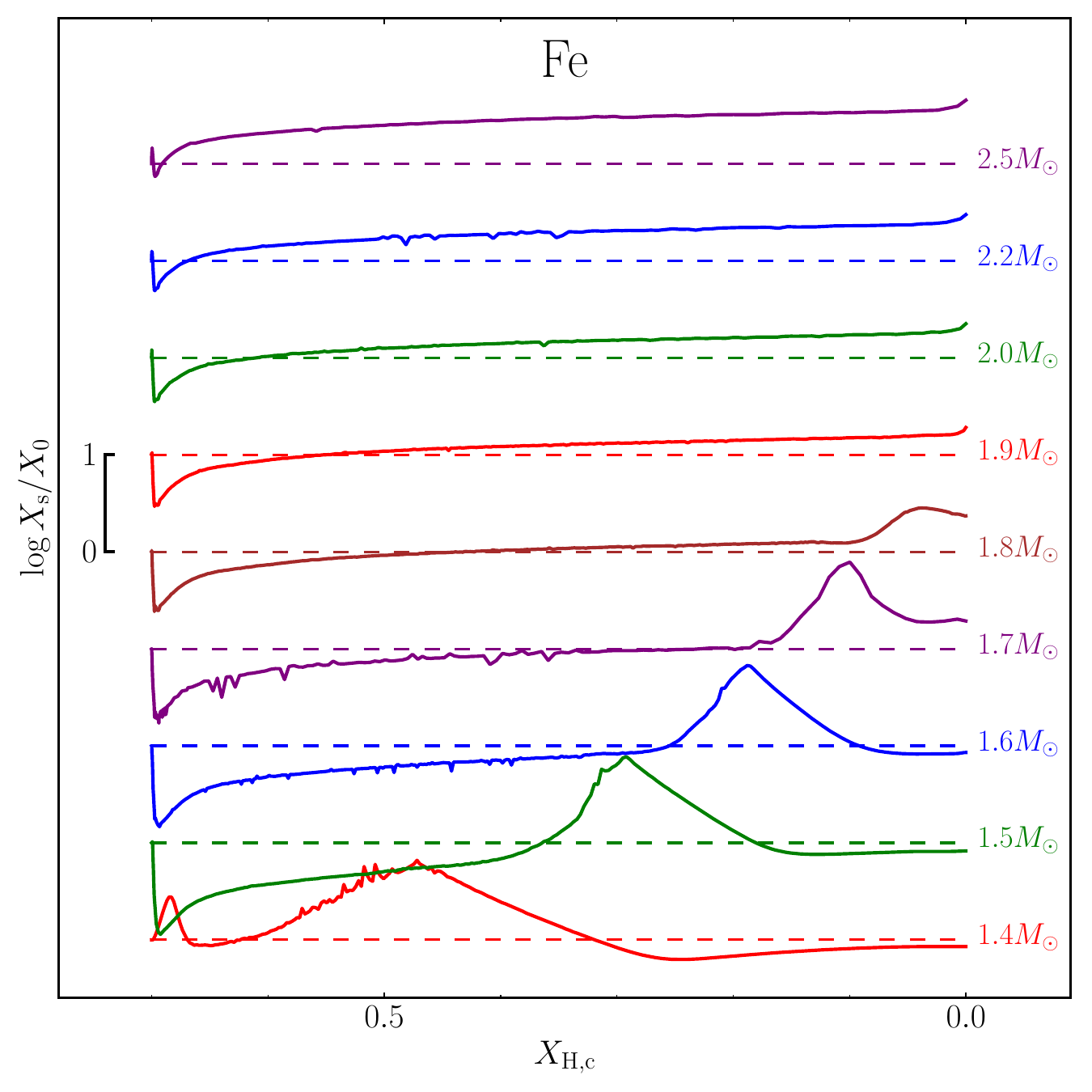}
 \caption{Evolutionary variations of the superficial iron abundance in stars with mass
          $1.4M_\odot\le M\le 2.5M_\odot$ and the mass loss rate $M=10^{-14}M_\odot/\textrm{yr}$.}
\label{fig7}
\end{figure}

\begin{figure}
 \centering
 \includegraphics[width=0.6\columnwidth,clip]{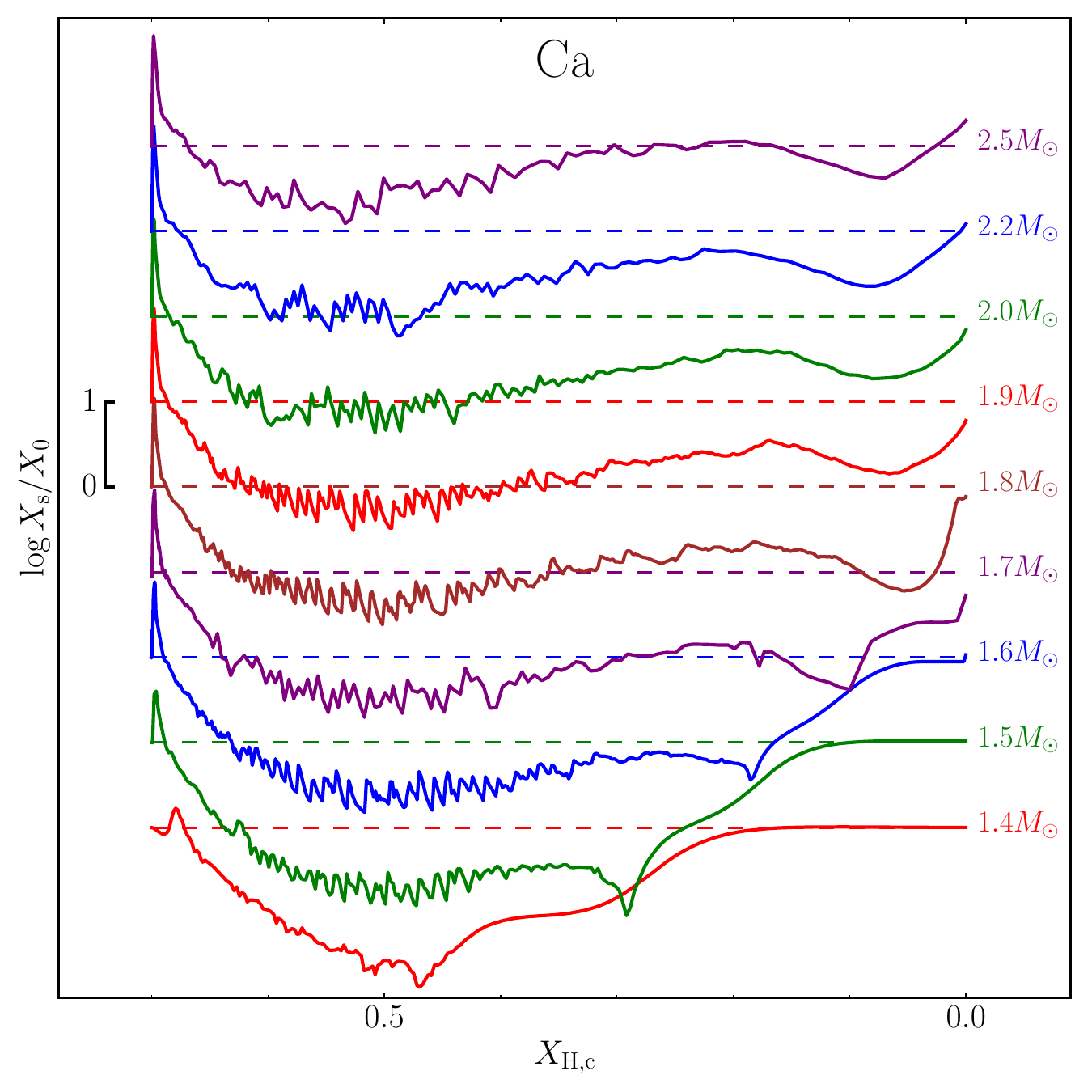}
 \caption{Same as Fig.~\ref{fig7} but for the superficial calcium abundance.}
\label{fig8}
\end{figure}

Figs.~\ref{fig7} and \ref{fig8} show the plots of evolutionary variations of the superficial
abundance of iron and calcium in stars with mass $1.4M_\odot\le M\le 2.5M_\odot$ and the mass loss
rate $M=10^{-14}M_\odot/\textrm{yr}$.
As seen in Fig.~\ref{fig7}, in stars with mass $M\ge 1.9M_\odot$ after the initial short--term
increase and decline the superficial iron abundance monotonously increases so that at the end
of the main sequence stage $X_\mathrm{s}$ is several times higher than its initial value $X_0$.
In stars with mass $M\le 1.8M_\odot$ the monotonous abundance growth is superimposed by variations
due to merge of the outer and intermediate convection zones.
Effect of a merging of convection zones is highest in stars with masses 1.5 and $1.6M_\odot$ where
the ranges of
evolutionary variations of the superficial abundance are as high as two orders of magnitude.

\begin{figure}
 \centering
 \includegraphics[width=0.6\columnwidth,clip]{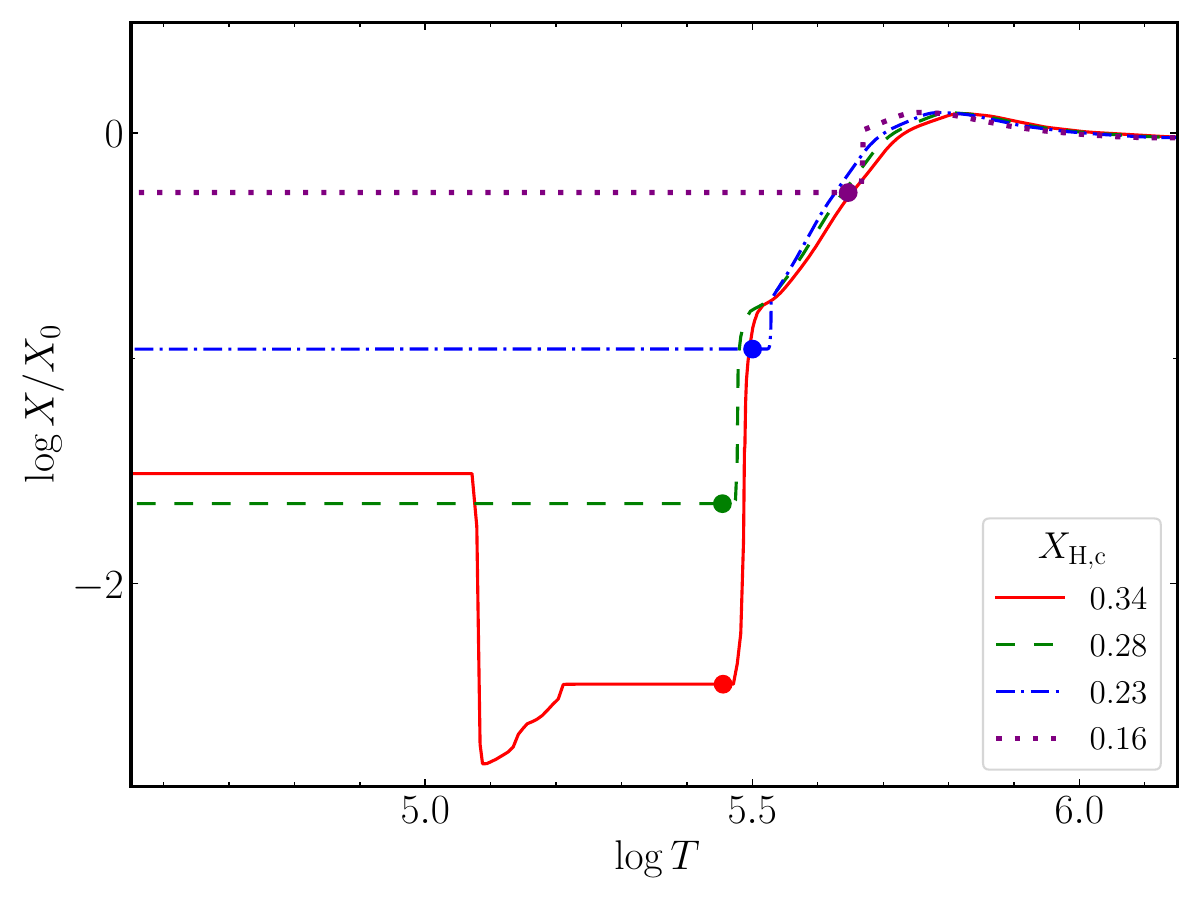}
 \caption{Radial distribution of calcium abundance in the four stellar models of the evolutionary
          sequence $M=1.5M_\odot$, $\dot M=10^{-14}M_\odot/\mathrm{yr}$
          at the central hydrogen abundance $X_\mathrm{H,c}=0.34$ (solid line), 0.28 (dashed line),
          0.23 (dash--dotted line) and 0.16 (dotted line). Circles in curves indicate location
          of the bottom of the outer convection zone.}
\label{fig9}
\end{figure}

Radiation pressure on calcium atoms is significantly smaller and therefore during the major part
of evolution time calcium atoms sink so that at the stellar surface appears deficiency of calcium.
Comparison of plots in Figs.~\ref{fig7} and \ref{fig8} for stars with mass $1.4M_\odot\le M\le 1.8M_\odot$
allows us to conclude that the merging of convection zones perceptibly affects the superficial calcium
abundance due to cease of its decrease and the following growth.
Initial monotonous decrease of superficial abundance of calcium is due to the fact that the layers
with temperature $T\approx 1.2\times 10^5\:\textrm{K}$ where calcium is most opaque
locate within the outer convection zone and the role of radiation pressure is canceled due to
convective mixing.
After the merging of convection zones the outer convection zone plunges into gas layers where
the number density of calcium atoms increases with decreasing radius so that
the superficial calcium abundance begins to growth.
This process is illustrated by plots in Fig.~\ref{fig9} for the four models of the evolutionary
sequence $M=1.5M_\odot$, $\dot M = 10^{-14}M_\odot/\textrm{yr}$.

\section{COMPARISON WITH OBSERVATIONS}

The Am stars most appropriate for comparison with results of our computations seem to be those
in the open clusters Praesepe and Hyades with the mean indices of metallicity
$[\textrm{Fe}/\textrm{H}]=0.16$ and $[\textrm{Fe}/\textrm{H}]=0.12$ (Netopil et al. 2022).
Assuming that the solar metallicity is $Z_\odot=0.014$ (Asplund et al. 2022) we obtain
that the metal abundances of Praesepe and Hyades are $Z=0.020$ and $Z=0.018$, respectively.
Observational estimates of the effective temperature $T_\mathrm{eff}$ and surface gravity
we took from (Hui--Bon--Hoa et al. 1997; Hui--Bon--Hoa and Alecian 1998;
Varenne and Monier 1999; Fossati et al. 2007; Gebran et al. 2010).
To determine the bolometric luminosity $L$ we used the trigonometric parallaxes from
the Gaia DR3 mission (Gaia Collaboration 2021), the bolometric corrections by
Creevey et al. (2022), the interstellar extinction maps by Lallement et al. (2022)
and the extinction law $R_V \times A_\lambda/A_v$ with $R_V=3.1$ (Mathis 1990).

The masses and ages of selected stars were determined by the isochrone method on the basis
of 11 evolutionary tracks of main sequence stars with masses $1.5M_\odot\le M\le 2.5M_\odot$
computed with taking into account effects of atomic diffusion.
It should be noted that effects of mass loss rates ranged from 0 to
$\dot M = 10^{-12}M_\odot/\textrm{yr}$ are negligible.

\begin{figure}
 \centering
 \includegraphics[width=0.7\columnwidth,clip]{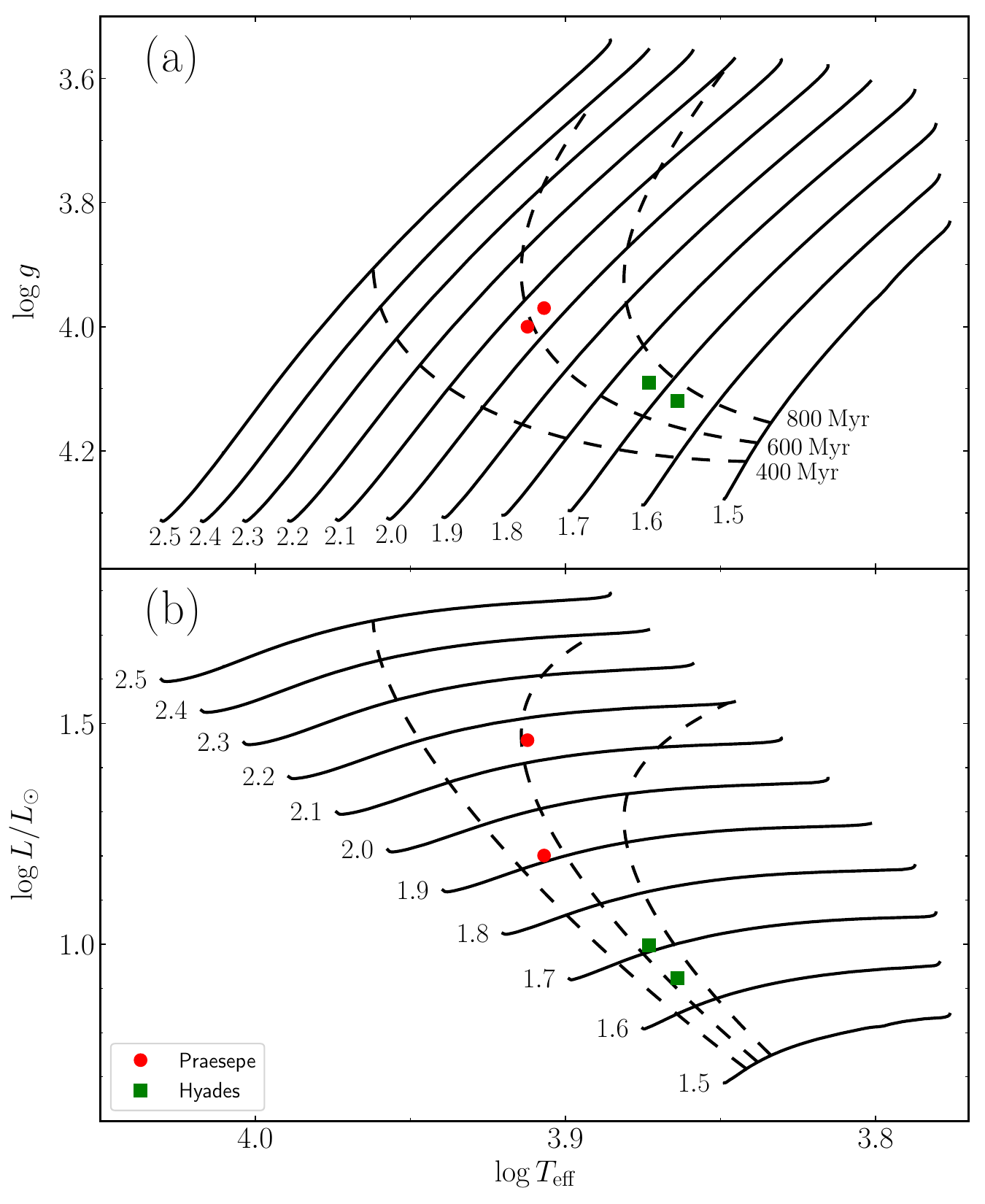}
 \caption{Evolutionary tracks (solid lines) and isohrones (dashed lines) on the Kiel (a) and
          Hertzsprung--Russel diagrams (b). Stellar masses and ages are shown at the initial points
          of the tracks and isochrones. Am stars of the open clusters are shown by circles
          (Praesepe) and squares (Hyades).}
\label{fig10}
\end{figure}

Employing the isochrone method we obtained for each selected star two couples of the mass and
age estimates.
The first pair $(M_1, t_\mathrm{ev,1})$ was evaluated with isochrones on the
Kiel diagram $\log T_\mathrm{eff} - \log g$, whereas the second pair
$(M_2, t_\mathrm{ev,2})$ with isochrones on the Hertzsprung--Russel diagram.
Finally we succeeded to select only four Am stars with small enough difference in
the couples of estimates $(M_1,M_2)$ and $(t_\mathrm{ev,1},t_\mathrm{ev,2})$.
These are (HD27628, HD28226) in Hyades with mass difference less than 1.5\% and
(HD73618, HD73730) in Praesepe with mass difference 5\% and 7\%, respectively.
Characteristics of selected stars are listed in Table~\ref{table1} whereas their location
on the Kiel and HR diagrams is shown in Fig.~\ref{fig10}.
The tracks in Fig.~\ref{fig10} correspond to monotonous decrease of the central hydrogen abundance
up to $X_\mathrm{H,c}\approx 0.04$ comprising $\approx 95\%$ of the main sequence lifetime.

\begin{figure}
 \centering
 \includegraphics[width=0.5\columnwidth,clip]{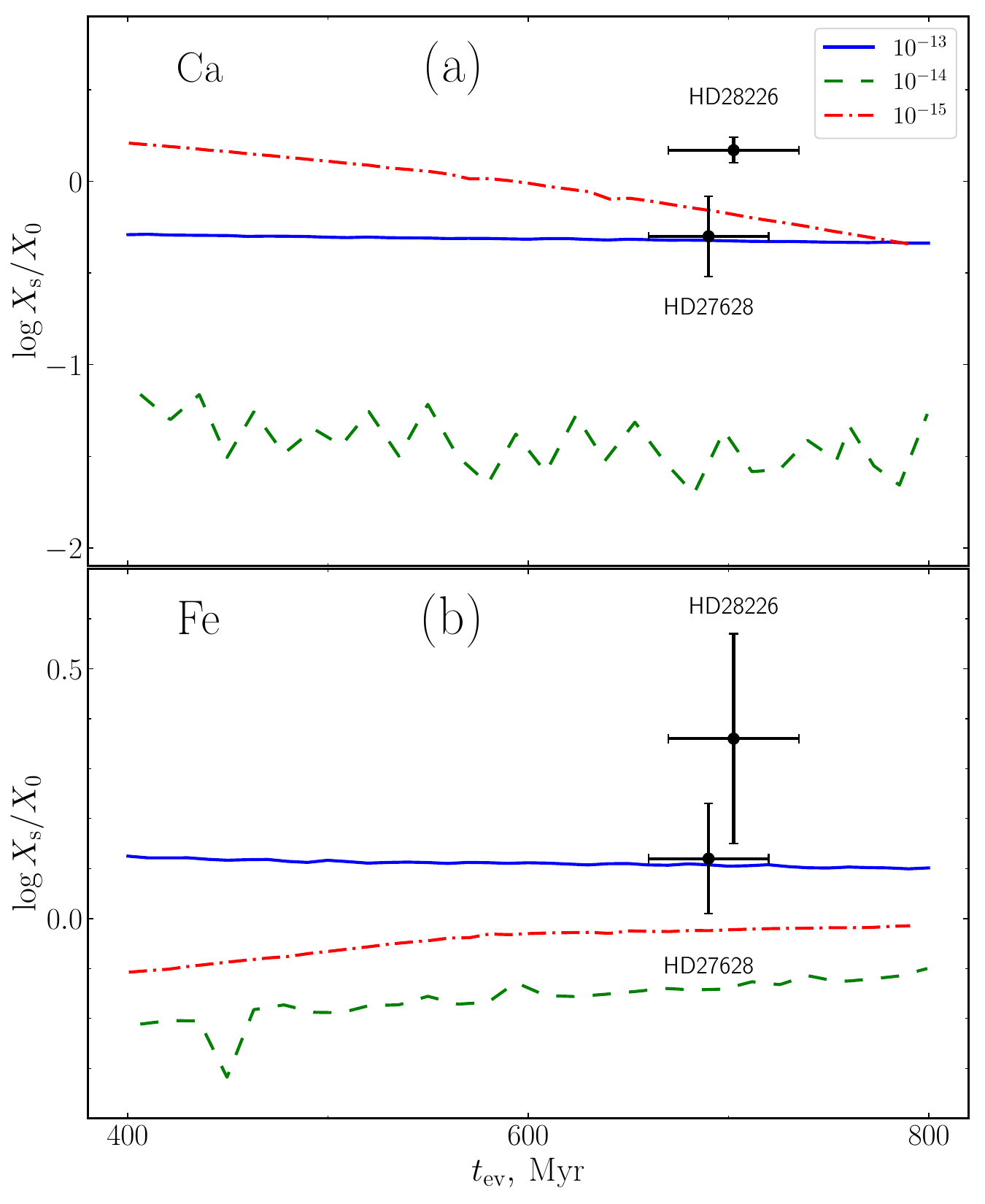}
 \caption{Variations of the superficial abundances of calcium (a) and iron (b) in the models
          of evolutionary sequences $M=1.7M_\odot$ and mass loss rates
          $\dot M=10^{-13}$, $10^{-14}$ и $10^{-15}M_\odot/\textrm{yr}$.
          Circles show the mean calcium and iron abundances in HD28226 and HD27628
          (Gebran et al. 2010).}
\label{fig11}
\end{figure}

\begin{figure}
 \centering
 \includegraphics[width=0.5\columnwidth,clip]{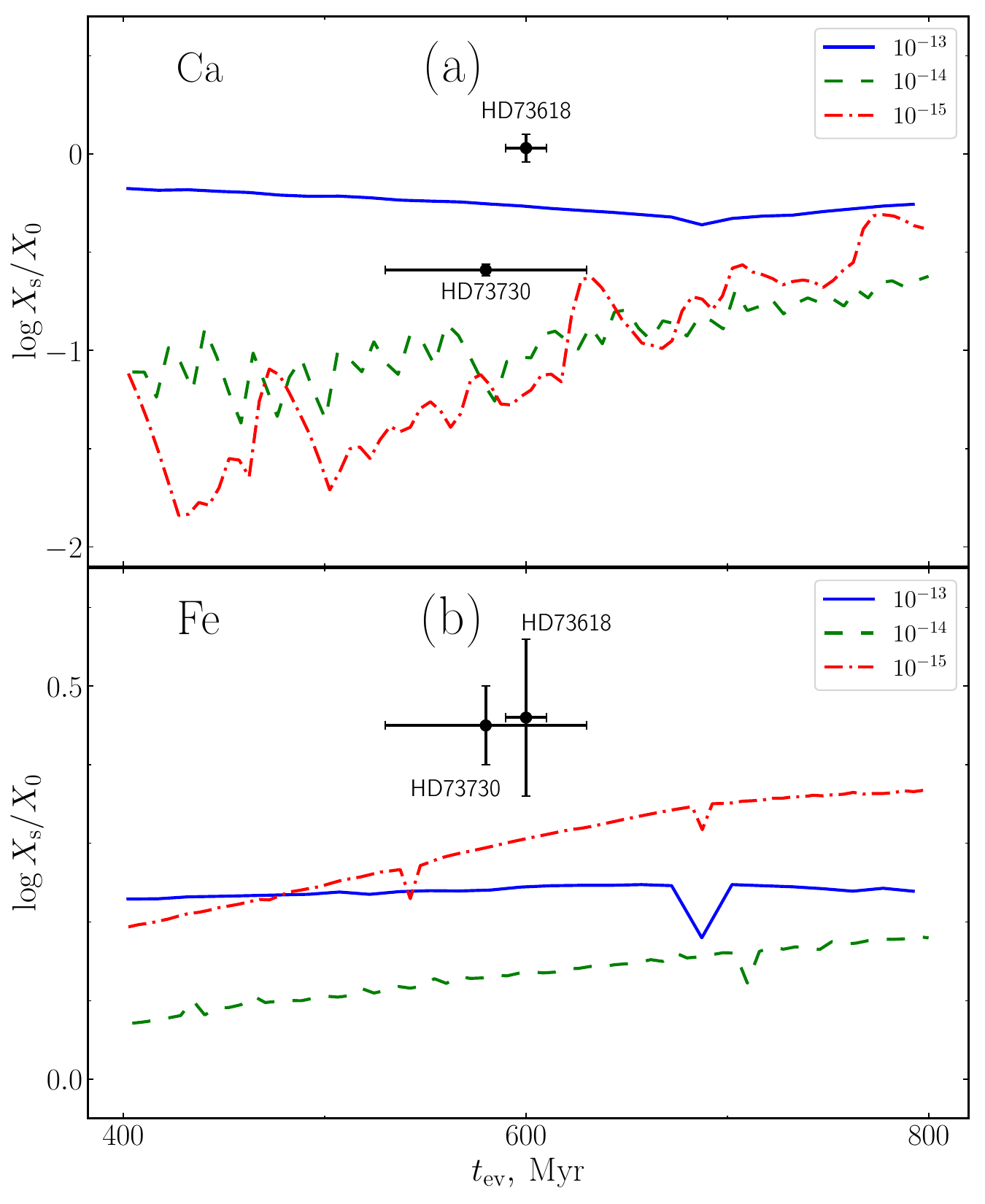}
 \caption{Same as Fig.~\ref{fig11} but for the models of evolutionary sequences with mass $M=2M_\odot$.
          Circles show the mean calcium and iron abundances in Am stars HD73618 and HD73730
          (Fossati et al. 2007).}
\label{fig12}
\end{figure}

Figs.~\ref{fig11} and \ref{fig12} show the evolutionary variations of the superficial abundance
of calcium and iron in the models of the evolutionary sequences with masses $M=1.7M_\odot$ and
$M=2.0M_\odot$.
As seen in Table~\ref{table1}, these plots can be used for comparison with observations of Am stars
HD28226 and HD27628 of Hyades (Gebran et al. 2010) as well as  HD73618 and HD73730 of Praesepe
(Fossati et al. 2007).
Insignificant difference between theoretical dependencies and observations demonstrates in favor
of the theory of stellar evolution.

\section{CONCLUSIONS}

The results of the present study allow us to conclude that the agreement between observations
and the diffusion models of stellar evolution can be obtained in the assumption of the
stellar wind in A stars of the main sequence.
The major argument in favor of the stellar wind is that the superficial abundance of chromium,
manganese and nickel significantly exceed observational abundances for low mass loss rates:
$\dot M < 10^{-14}M_\odot/\textrm{yr}$.
On the other hand, the intermediate convection zone does not appear in the models with
mass loss rate $\dot M\gtrsim 10^{-13}M_\odot/\textrm{yr}$ so that the superficial
abundances of iron and calcium vary within narrow range than observed chemical
anomalies in Am stars.

For mass loss rates $10^{-14}M_\odot/\textrm{yr}\lesssim\dot M\lesssim 10^{-13}M_\odot/\textrm{yr}$
results of calculations reveal agreement with observations.
Therefore, we conclude that Am stars are the nonmagnetic slowly rotating stars where anomalies
of calcium and iron appear due to evolutionary variations of the thickness of the convection zones
as well as diffusion elemental separation connected with gravitational settling and radiative
expulsion in the zones of radiative transfer.
Our theoretical mass loss rate estimates agree with that
$\dot M\sim 3\times 10^{-14}M_\odot/\textrm{yr}$
of Vick et al. (2010) and Hui--Bon--Hoa et al. (2022).

A major difficulty encountered in comparison of the theory with observations is due to
insufficiently high precision in the determination of the fundamental parameters of Am stars.
One of the ways to overcome this difficulty is to expand the list of Am stars with more
close mass and age estimates obtained by the isochrone method on the Kiel and HR diagrams.

\section*{REFERENCES}

\begin{enumerate}

 \item M.~Asplund, N.~Grevesse, A.J.~Sauval, P.~Scott,
       ``The chemical composition of the Sun'',
       Ann. Rev. Astron. Astrophys. \textbf{47}, 481 (2009).

 \item P.~Bertin, H.J.G.L.M.~Lamers H.J.G.L.M., A.~Vidal--Madjar, R.~Ferlet, and R.~Lallement,
       ``HST--GHRS observations of Sirius A. III. Detection of a stellar wind from Sirius A'',
       Astron. Astrophys. \textbf{302}, 899 (1995).

 \item E. B\"ohm-Vitense,
       ``\"Uber die Wasserstoffkonvektionszone in Sternen verschie\-dener Effektivtemperaturen und Leuchtkr\"afte'',
       Zeitschrift f{\"u}r Astrophys. \textbf{46}, 108 (1958).

 \item J.M.~Burgers, \textit{Flow equations for composite gases} (Academic Press, New York and London, 1969).

 \item H.~Caliskan and S.J.~Adelman,
       ``Elemental abundance analyses with DAO spectrograms - XVII. The superficially normal early A stars 2 Lyncis,
       omicron Ursa Majoris and phi Aquilae'',
       MNRAS \textbf{288}, 501 (1997).

 \item B.~Campilho, M.~Deal, and D.~Bossini,
       ``Atomic diffusion in solar-like stars with MESA. Comparison with the Montreal/Montpellier and
       CESTAM stellar evolution codes'',
       Astron. Astrophys. \textbf{659}, A162 (2022).

 \item A.~Cowley, C.~Cowley, M.~Jaschek, and C.~Jaschek,
       ``A study of the bright A stars. I. A catalogue of spectral classifications'',
       Astron.J. \textbf{74}, 375 (1969).

 \item P.S.~Conti,
       ``The metallic--line stars'',
       Publ. Astron. Soc. Pacific \textbf{82}, 781 (1970).

 \item O.~Creevey, A.~Vallenaro, and A.~Brown,
       ``A golden sample of astrophysical parameters'',
       EAS2022, European Astronomical Society Annual Meeting, 1449 (2022).

 \item M.~Deal, M.~J.~Goupil, and J.P.~Marques,
       ``Chemical mixing in low mass stars. I. Rotation against atomic diffusion including radiative acceleration'',
       Astron. Astrophys. \textbf{633}, A23 (2020).

 \item T.~Dumont, C.~Charbonnel, and A.~Palacios,
       ``Lithium depletion and angular momentum transport in F-type and G-type stars in Galactic open clusters'',
       Astron. Astrophys. \textbf{654}, A46 (2021).

 \item L.~Fossati, S.~Bagnulo, R.~Monier, S.A.~Khan, O.~Kochukhov, J.~Landstreet, G.~Wade, and W.~Weiss,
       ``Late stages of the evolution of A--type stars on the main sequence: comparison between observed chemical
       abundances and diffusion models for 8 Am stars of the Praesepe cluster'',
       Astron. Astrophys. \textbf{476}, 911 (2007).

 \item Gaia Collaboration (Gaia Collaboration),
       ``Gaia Early Data Release 3. Summary of the contents and survey properties'',
        Astron. Astrophys. \textbf{649}, A1 (2021).

 \item M.~Gebran, M.~Vick, R.~Monier, and L.~Fossati,
       ``Chemical composition of A and F dwarfs members of the Hyades open cluster'',
       Astron. Astrophys. \textbf{523}, A71 (2010).

 \item S.~Ghazaryan, G.~Alecian, and A.A.~Hakobyan,
       ``New catalogue of chemically peculiar stars, and statistical analysis'',
       MNRAS, \textbf{480}, 2953 (2018).

 \item N.~Grevesse and A.J.~Sauval,
       ``Standard Solar Composition'',
       Space Science Rev. \textbf{85}, 161 (1998).

 \item F. Herwig,
       ``The evolution of AGB stars with convective overshoot'',
       Astron. Astrophys. \textbf{360}, 952 (2000).

 \item H.~Hu, C.A.~Tout, E.~Glebbeek, and M.--A.~Dupret,
       ``Slowing down atomic diffusion in subdwarf B stars: mass loss or turbulence?'',
       MNRAS \textbf{418}, 195 (2011).

 \item A.~Hui--Bon--Hoa, C.~Burkhart, and G.~Alecian,
       ``Metal abundances of A--type stars in three galactic clusters'',
       Astron. Astrophys. \textbf{323}, 901 (1997).

 \item A.~Hui--Bon--Hoa,
       ``Radiative acceleration calculation methods and abun\-dance anomalies in Am stars'',
       Astron. Astrophys. \textbf{691}, A266 (2024).

 \item A.~Hui--Bon--Hoa and G.~Alecian,
       ``Metal abundances of A--type stars in galactic clusters. II. Pleiades, Coma Berenices, Hyades, and Praesepe'',
       Astron. Astrophys. \textbf{332}, 224 (1998).

 \item A.~Hui--Bon--Hoa, G.~Alecian, and F.~LeBlanc,
       ``Modelling of the scandium abundance evolution in AmFm stars'',
       Astron. Astrophys. \textbf{668}, A6 (2022).

 \item C.A.~Iglesias, F.J.~Rogers, and B.G.~Wilson,
       ``Reexamination of the metal contribution to astrophysical opacity'',
       Astrophys. J. \textbf{322}, L45 (1987).

 \item R.~Lallement, J.L.~Vergely, C.~Babusiaux, and N.L.J.~Cox,
       ``Updated Gaia-2MASS 3D maps of Galactic interstellar dust'',
       Astron. Astrophys. \textbf{661}, A147 (2022).

 \item N.~Langer, M.F.~El~Eid, and K.J.~Fricke,
       ``Evolution of massive stars with semiconvective diffusion'',
       Astron. Astrophys. \textbf{145}, 179 (1985).

 \item L.I.~Mashonkina and Yu.A. Fadeyev,
       ``Revision of the calcium and scandium abundances in Am stars based on non-LTE calculations
       and comparison with diffusion stellar evolution models'',
       Astron. Lett. \textbf{50}, 373 (2024)].

 \item J.~Mathis,
       ``Interstellar dust and extinction'',
       Ann. Rev. of Astron. and Astrophys. \textbf{28}, 37 (1990).

 \item G.~Michaud,
       ``Diffusion processes in peculiar A stars'',
       Astrophys. J. \textbf{160}, 641 (1970).

 \item G.~Michaud, Y.~Charland, S.~Vauclair, and G.~Vauclair,
       ``Diffusion in main--sequence stars: radiation forces, time scales, anomalies'',
       Astrophys. J. \textbf{210}, 447 (1976).

 \item G.~Michaud, J.~Richer, and M.~Vick,
       ``Sirius A: turbulence or mass loss?'',
       Astron. Astrophys. \textbf{534}, A18 (2011).

 \item N.~Moedas, M.~Deal, D.~Bossini, and B.~Campilho,
       ``Atomic diffusion and turbulent mixing in solar-like stars: Impact on the fundamental properties of FG--type stars'',
       Astron. Astrophys. \textbf{666}, A43 (2022).

 \item M.~Netopil, \.{I}.A.~Oralhan, H.~\c{C}akmak, R.~Michel, Y.~Karata\c{s},
       ``The Galactic metallicity gradient shown by open clusters in the light of radial migration'',
       MNRAS, \textbf{509}, 421 (2022).

 \item B. Paxton, J. Schwab, E.B. Bauer, L. Bildsten, S. Blinnikov, P. Duffell, R. Farmer,
       J.A. Goldberg, P. Marchant, E. Sorokina, A. Thoul, R.H.D. Townsend, and F.X. Timmes,
       ``Modules for Experiments in Stellar Astrophysics (MESA): Convective Boundaries, Element Diffusion,
       and Massive Star Explosions'',
       Astrophys. J. Suppl. Ser. \textbf{234}, 34 (2018).

 \item G.W.~Preston,
       ``The chemically peculiar stars of the upper main sequence'',
       Ann. Rev. Astron. Astrophys. \textbf{12}, 257 (1974).

 \item P.~Renson and J.~Manfroid,
       ``Catalogue of Ap, HgMn and Am stars'',
       Astron. Astrophys. \textbf{498}, 961 (2009).

 \item J.~Richer, G.~Michaud, and C.R.~Proffitt,
       ``Helium gravitational settling in the envelopes of evolving main--sequence A and F stars'',
       Astrophys. J. Suppl. Ser. \textbf{82}, 329 (1992).

 \item J.~Richer and G.~Michaud,
       ``Diffusion of lithium and beryllium in evolving, nonrotating main--sequence A and F stars'',
       Astrophys. J. \textbf{416}, 312 (1993).

 \item J.~Richer, G.~Michaud, and S.~Turcotte,
       ``The evolution of AMFM stars, abundance anomalies, and turbulent transport'',
       Astrophys. J. \textbf{529}, 338 (2000).

 \item N.G.~Roman, W.W.~Morgan, and O.J.~Eggen,
       ``The classification of the metallic--line stars'',
       Astrophys. J. \textbf{107}, 107 (1948).

 \item M.J.~Seaton,
       ``Opacity Project data on CD for mean opacities and radiative accelerations'',
       MNRAS \textbf{362}, L1 (2005).

 \item M.J.~Seaton and N.R.~Badnell,
       ``A comparison of Rosseland-mean opacities from OP and OPAL'',
       MNRAS \textbf{354}, 457 (2004).

 \item A.~Slettebak,
       ``A catalogue of the brighter metallic--line stars'',
       Astrophys. J. \textbf{109}, 547 (1949).

 \item A.A.~Thoul, J.N.~Bahcall, and A.~Loeb,
       ``Element diffusion in the solar interior'',
       Astrophys. J. \textbf{421}, 828 (1994).

 \item S.~Turcotte, J.~Richer, and G.~Michaud,
       ``Consistent evolution of F stars: diffusion, radiative accelerations, and abundance anomalies'',
       Astrophys. J. \textbf{504}, 559 (1998).

 \item O.~Varenne and R.~Monier,
       ``Chemical abundances of A and F-type stars: the Hyades open cluster'',
       Astron. Astrophys. \textbf{351}, 247 (1999).

 \item M.~Vick, G.~Michaud, J.~Richer, and O.~Richard,
       ``AmFm and lithium gap stars. Stellar evolution models with mass loss'',
       Astron. Astrophys. \textbf{521}, A62 (2010).

 \item G.~Vauclair, S.~Vauclair, and G.~Michaud,
       ``Abundance anomalies in main--sequence stars: competition between diffusion processes and turbulent motions'',
       Astrophys. J. \textbf{223}, 920 (1978).

\end{enumerate}

\begin{table}
\caption{Characteristics of selected Am stars}
\label{table1}
\begin{center}
\begin{tabular}{cccccccc}
\hline
 HD & $T_\mathrm{eff}$ & $\log g$ & $\log L/L_\odot$ & $M_1/M_\odot$ & $t_\mathrm{ev,1}$ & $M_2/M_\odot$ & $t_\mathrm{ev,2}$ \\
\hline
27628 &  7310 &  4.12 & 0.923 & 1.67 & 720 & 1.65 & 660 \\
28226 &  7465 &  4.09 & 0.998 & 1.73 & 735 & 1.71 & 670 \\
73618 &  8170 &  4.00 & 1.462 & 2.05 & 590 & 2.15 & 610 \\
73730 &  8070 &  3.97 & 1.201 & 2.05 & 630 & 1.91 & 530 \\
\hline
\end{tabular}
\end{center}
\end{table}

\newpage

\end{document}